\documentclass[12pt]{article}
\usepackage[a4paper,margin=1in]{geometry}
\usepackage{graphicx}
\usepackage{hyperref}
\usepackage{amsmath, amssymb}
\usepackage{natbib}
\usepackage{placeins}
\usepackage{xcolor}
\usepackage{mhchem}
\usepackage{float}
\usepackage[utf8]{inputenc}

\DeclareUnicodeCharacter{2713}{\checkmark}

\title{Materials Informatics Across the Length Scales}
\author{
Jamal Abdul Nasir\textsuperscript{5},
Hamide Kavak\textsuperscript{1}, O\u{g}uzhan Der\textsuperscript{2}, Ali Ercetin\textsuperscript{2},\\
Amila Akagic\textsuperscript{3}, Jesper Friis\textsuperscript{4}, Francesca L. Bleken\textsuperscript{4},
Andrea Lorenzoni\textsuperscript{6},\\
Francesco Mercuri\textsuperscript{6}, Scott M. Woodley\textsuperscript{5}, Keith T. Butler\textsuperscript{5}\\
\small \textsuperscript{1} Department of Physics, \c{C}ukurova University, Adana, 01330, T\"{u}rkiye\\
\small \textsuperscript{2} Band\i rma Onyedi Eyl\u{u}l University, T\"{u}rkiye\\
\small \textsuperscript{3} University of Sarajevo, Bosnia and Herzegovina\\
\small \textsuperscript{4} SINTEF, Norway\\
\small \textsuperscript{5} Department of Chemistry, University College London, WC1E 6BS, UK\\
\small \textsuperscript{6} Consiglio Nazionale delle Ricerche (CNR), Italy
}

\begin{document}
\maketitle


\begin{abstract}Materials informatics is increasingly used to support modelling, analysis and design across the length scales of materials science, from atomistic simulations to microstructural characterisation and continuum descriptions. Despite rapid progress, the reliability and transferability of these approaches vary strongly with scale. Here we survey data-driven methods at the nanoscale, mesoscale, and micro-to-continuum levels, highlighting established capabilities as well as unresolved challenges. Machine-learning interatomic potentials, mesoscale surrogate and operator-learning models, and learning-based analysis of experimental microstructures are discussed, with emphasis on data quality, uncertainty, interpretability, and cross-scale consistency. We further examine the role of data standards, ontologies, and emerging tools, such as autonomous laboratories, where they directly affect multiscale workflows. This perspective clarifies what can be considered reliable today and identifies key obstacles to the broader integration of materials informatics across scales. 
\end{abstract}

\section{Introduction}
Materials science seeks to map a material’s properties across the full space of experimentally accessible conditions, spanning length scales from atomic to component scales and demanding explanations that connect phenomena across these scales \cite{noack2020autonomous}. Materials informatics has emerged to meet this challenge, integrating heterogeneous data and models from electronic and atomic  descriptions through nanoscale building blocks and mesoscale microstructures to micro-to-continuum behaviour to uncover structures that can augment or even replace conventional analyses \cite{rajan2005materials}, \cite{chen2020materials}, \cite{oommen2022learning}. In recent years, data-driven methods and machine learning (ML) have become pervasive across the field, enabling the discovery of nonlinear structure–property relationships that are difficult to capture with traditional paradigms  \cite{ramprasad2017machine, PEIVASTE2025119419}. A notable example is the rise of symmetry-equivariant interatomic potentials, which deliver forces with near first-principles fidelity over long molecular-dynamics trajectories, providing reliable atomistic inputs for higher-scale models and thereby strengthening multiscale linkages \cite{batzner20223}. 

The multiscale nature of materials science is mirrored by the diverse roles that materials informatics now plays across the field \cite{butler2018machine}. 
Data-driven techniques are applied to accelerate simulations and predict materials properties via high-throughput computational screening and  surrogate models that reduce the need for expensive first-principles calculations \cite{jain2013commentary} \cite{schmidt2019recent}, automate experiments and close the design–make–measure loop through adaptive sampling and uncertainty-guided active learning \cite{lookman2019active}, analyse complex and heterogeneous datasets by leveraging ML for feature extraction and pattern recognition across structural, spectroscopic, and performance data \cite{schmidt2019recent}, extract knowledge from the literature using informatics tools drawn from materials databases and semantic mining frameworks, and integrate multiscale computational workflows that bridge atomistic to continuum models in predictive design pipelines. \cite{LePiane2024}  At the computational level, ML is increasingly used to accelerate atomistic and electronic-structure simulations, most prominently through machine-learning interatomic potentials (MLIPs) and surrogate models that reproduce first-principles accuracy at greatly reduced cost \cite{behler2007generalized}, \cite{bartok2010gaussian}. In parallel, informatics-driven optimisation and active-learning strategies have enabled automated and closed-loop experimentation, including autonomous synthesis and characterisation platforms for materials discovery \cite{macleod2020self}. Machine learning has also become a central tool for quantitative analysis of high-dimensional experimental data, such as microstructural images, diffraction patterns and spectroscopy, enabling objective and high-throughput interpretation beyond manual approaches \cite{azimi2018advanced}. Beyond numerical data, natural-language processing and text-mining methods are increasingly applied to extract structure–property relationships and trends from the scientific literature and databases \cite{tshitoyan2019unsupervised}. Across these applications, predictive models are most often used to prioritise candidates, guide experiments and connect heterogeneous data sources, thereby shortening design–make–measure cycles rather than replacing physical modelling or experimentation outright \cite{jain2013commentary}. 

A recent example used high-throughput computational techniques to screen thousands of potential alloy compositions for hydrogen storage applications~\cite{altintas2023shoulders}. This accelerated the identification of promising materials by predicting their hydrogen absorption and release properties with high accuracy, reducing the need for extensive experimental testing. Another recent study developed ML models to predict the properties of materials based on composition and structure data~\cite{chibani2020machine}. Enabled rapid identification of materials with desirable properties (e.g., thermal conductivity, mechanical strength) without the need for costly and time-consuming experiments.
ML/DL (Deep Learning) is also being applied in advanced characterisation of materials. Researchers recently implemented robotic systems for the automated synthesis and characterization of perovskite materials~\cite{sun2019accelerated}. This enhanced the speed and efficiency of discovering new perovskite materials with potential applications in photovoltaics and optoelectronics. 

Open, Python-centred ecosystems now underpin the field, lowering barriers and improving reproducibility: general ML libraries (scikit-learn, PyTorch, TensorFlow, JAX)\cite{pedregosa2011scikit}\cite{paszke2019pytorch}, domain toolkits (ASE, pymatgen, AiiDA)\cite{larsen2017atomic} and major data platforms (Materials Project, NOMAD, JARVIS, AFLOW)\cite{jain2013commentary} sit alongside interoperability standards such as OPTIMADE\cite{andersen2021optimade}. Together they enable end-to-end, cross-scale workflows, from atomistic modelling through micro- to continuum analysis and support both representation-learning models for images, spectra, diffraction and probabilistic approaches for design and uncertainty, notably Gaussian-process regression for active-learning loops. Emerging large-language-model tools promise to streamline literature mining, workflow automation and code generation, although their impact on discovery remains to be established. 

MLIPs now span a wide design space system dimensionality, interaction range, and the degree to which physics is built in and are often organised into a four-generation scheme \cite{ko2021general}\cite{behler2007generalized}. First-generation models target low-dimensional problems (typically small molecules), delivering accurate potential-energy surfaces but limited transferability. Second-generation models exploit locality by decomposing the total energy into atomic contributions, which enables simulations of systems with thousands of atoms while retaining near first-principles accuracy. Third-generation models relax the pure-locality assumption by adding explicit long-range terms (for example, Coulombic electrostatics and dispersion) without ad-hoc truncation, yet they still infer charges from local environments. Fourth-generation approaches go further by coupling the potential to a global charge/electrostatic description (e.g., charge-equilibration or self-consistent field updates), allowing genuinely nonlocal phenomena such as long-range charge transfer to be captured\cite{behler2021four}.

While there are scale- and domain-specific issues, several cross-cutting challenges recur throughout this Review. A first cluster concerns data: despite a proliferation of high-quality, simulation-generated databases, standardised experimental datasets with rich and reliable metadata remain relatively scarce \cite{draxl2018nomad}. Data scarcity, inconsistency, errors and heterogeneous formats all limit the effective training and benchmarking of ML models; providing open, well-annotated and interoperable data is therefore essential for progress. A second cluster relates to methods and infrastructure. The computational cost of state-of-the-art models can be substantial, making it difficult to scale to large design spaces or long time horizons, and integrating models across length scales in a robust multiscale framework remains a grand challenge\cite{peng2020multiscale}. Finally, limited interpretability and incomplete uncertainty quantification continue to hinder trust in ML predictions for high-stakes decisions\cite{tavazza2021uncertainty}, and difficulties in sharing data and models across incompatible formats underscore the need for broader adoption of common standards and platforms such as FAIR data infrastructures\cite{andersen2021optimade, LePiane2024}. 

In this Review, we examine how materials informatics is being applied across different length scales in materials science, providing both an integrated perspective on the field and a basis for identifying shared challenges and opportunities for cross-disciplinary exchange. The sections that follow consider machine-learning interatomic potentials at the nanoscale, data-driven and surrogate modelling approaches at the mesoscale, and learning methods that bridge microstructural and continuum descriptions. Throughout, we emphasise what can be considered reliable at present and where transferability across scales remains limited. We also discuss data standards and emerging capabilities such as autonomous experimentation and language-model-based tools, where they have a direct impact on cross-scale workflows. The Review concludes with recommendations aimed at mitigating the challenges identified. 

\section{State of the art and challenges across the length-scales}
Developing methods and theories that both explain experimental observations and guide new experiments remains central to materials science, yet predicting properties consistently across length scales is still challenging. ML is now a practical engine for linking materials data to predictions across scales. Rather than replacing domain expertise, ML reduces the manual burden of hypothesis screening by learning patterns directly from data, whether structured state variables (temperature, pressure, composition) or raw characterisation outputs (spectra, diffraction, microscopy images)\cite{botifoll2022machine}. Modern models accept both hand-crafted descriptors and minimally processed inputs, and they integrate cleanly with established simulation and analysis workflows. Reported properties span atomistic energies \cite{behler2016perspective}\cite{li2015molecular} and crystal structure tasks \cite{schmidt2019recent}, microscopic strain mapping \cite{yang2019establishing}, and macroscopic responses such as compressive strength \cite{gallagher2020predicting}, electronic conductivity \cite{zhang2020dramatically}, and thermal stability \cite{lu2020interpretable}. Used with appropriate validation and uncertainty estimation, these tools provide a consistent way to move information between the atomic, nano-, meso-, and micro-to-continuum regimes, setting up the scale-specific methods surveyed in the subsections that follow. For instance, MLIPs can extend the spatial and temporal scales of molecular-dynamics simulations while retaining near first-principles accuracy, helping to transmit atomic-scale insight into mesoscale and device-level models. The subsections that follow detail methods and exemplars at the nano-, meso-, and micro-to-continuum levels and how they integrate across scales. 

\subsection{Nanoscale learning}
The nanoscale refers not only to characteristic dimensions on the order of nanometres, but more importantly to regimes in which surfaces, interfaces, finite size effects, and symmetry breaking give rise to physical and chemical behaviour that cannot be captured by bulk or continuum descriptions \cite{jia2021machine}. ML models can identify patterns and correlations within these datasets that may not be apparent through conventional analysis methods, which allows for the discovery of new materials with specific desired properties by predicting how variations in composition, structure, and processing conditions will affect the final material. For instance, Champa-Bujaico \textit{et al}.~\cite{champa2024optimization} used a combination of experimental and ML techniques to optimise the mechanical properties of multiscale hybrid polymer nanocomposites. Similarly, Xiao \textit{et al}.~\cite{xiao2020machine} discuss how materials informatics serves as a "super-accelerator" in the context of nanoarchitectonics, a concept that integrates nanotechnology and materials science to create functional materials with tailored properties. Their study highlights the use of ML models to predict the synthesis conditions for nanoporous materials, which are essential for applications in catalysis and energy storage. By employing active learning and Bayesian optimisation, the researchers were able to significantly reduce the time required to identify optimal material compositions and synthesis conditions, thus accelerating the material development process. 

\begin{figure}[H]
  \centering
  \includegraphics[width=\linewidth]{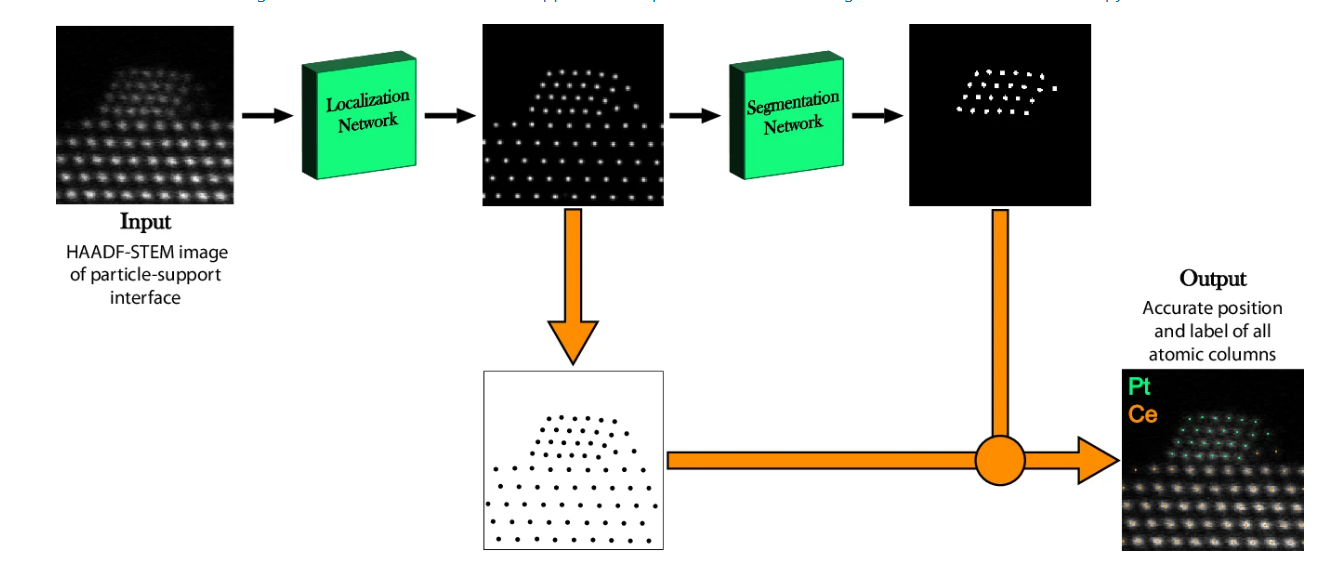}
  \caption{ \textbf{Model overview.} Deep-learning workflow for automated localisation and classification of atomic columns in HAADF-STEM images of supported nanoparticles. Two U-Net-based networks are used to achieve sub-pixel column localisation and subsequent particle–support segmentation, enabling quantitative atomic-scale analysis under noisy, low-dose imaging conditions. Reproduced with permission from Ref. \cite{eliasson2024localization}. }
  \label{fig:Figure1}
\end{figure}

Recent deep-learning approaches have transformed atomic-resolution Scanning transmission electron microscopy (STEM) imaging from a qualitative characterisation tool into a quantitative, data-driven source of nanoscale structural descriptors as shown in Figure~\ref{fig:Figure1}, \cite{eliasson2024localization}. Such learning-based microscopy pipelines exemplify a broader shift in nanoscale materials informatics: experimental images are no longer treated as qualitative visual evidence, but as high-dimensional data sources from which quantitative, atomically resolved descriptors can be extracted automatically. The resulting atomic coordinates and labels provide a structured representation that can be directly coupled to downstream analyses, including strain mapping, defect tracking, statistical analysis of dynamic processes, and comparison with atomistic simulations. Importantly, these methods also enable time-resolved studies of nanoparticle dynamics at sub-second temporal resolution, thereby opening new opportunities to connect experimental observations with data-driven models of nanoscale structure–property relationships.

\begin{figure}[H]
  \centering
  \includegraphics[width=\linewidth]{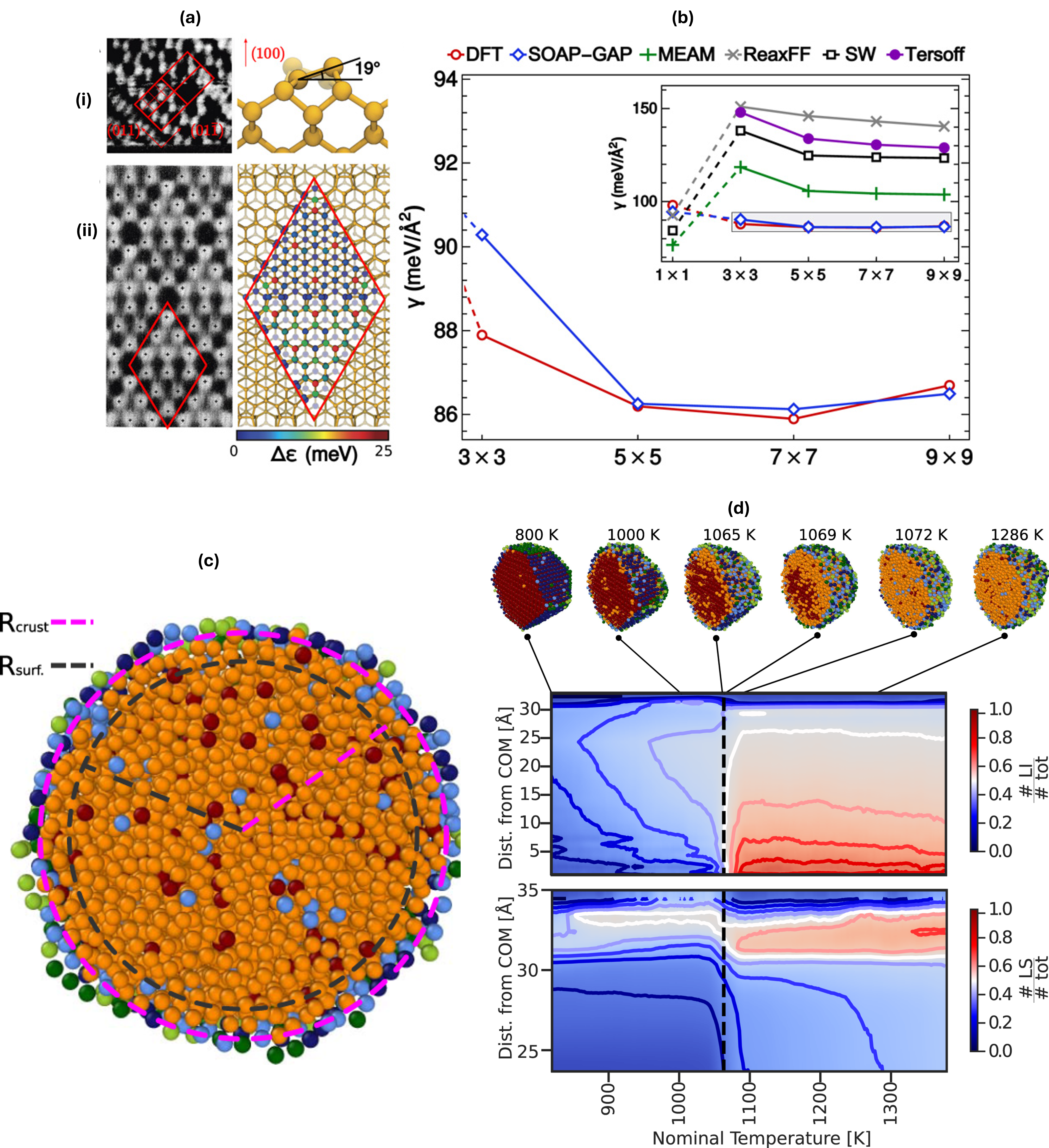}
  \caption{\textbf{MLIP predictions for silicon surface reconstructions and Au nanoparticle melting.}
  \textbf{a}, Si(111)\,(7$\times$7) reconstruction and Si(100) dimer tilt. Left: scanning tunnelling microscopy (STM). Right: SOAP--GAP results: (i) Si(111) structure relaxed with SOAP--GAP and coloured by per-atom predicted local energy error when trained without adatoms; (ii) Jahn--Teller--induced $\sim$19$^\circ$ dimer tilt on Si(100). Reproduced with permission from Ref.~\cite{bartok2017machine} (CC BY 4.0). SOAP--GAP captures the tilt whereas empirical models miss this feature.
  \textbf{b}, Energy ordering of Si(111) reconstructions: SOAP--GAP correctly identifies the 7$\times$7 DAS ground state, in contrast to empirical force fields.
  \textbf{c}, Au$_{6266}$ nanoparticle with $R_{\mathrm{crust}}$ (pink) and $R_{\mathrm{surf}}$ (grey); atoms coloured by unsupervised cluster class.
  \textbf{d}, Spatial--thermal distribution of liquid-like (\#LI) and surface-like (\#LS) local environments in Au$_{6266}$ from MLIP molecular dynamics: heat maps show average fractions versus radial distance from the centre of mass (vertical) and temperature (horizontal); bold isosurfaces from 0 to 1 in steps of 0.1; dashed line marks $T_{\mathrm{melt}}$. Panels \textbf{c},\textbf{d} reproduced with permission from Ref.~\cite{zeni2021data}.}
  \label{fig:fig1}
\end{figure}

One of the significant advances in materials informatics is the prediction of mechanical properties in nanocomposites.  MLIPs provide a complementary route to learning nanoscale structure–property relationships directly from quantum-mechanical data such as density functional theory (DFT) \cite{mortazavi2023atomistic}. In recent work on supported metal nanoparticles, Maxson and Szilvási employed an MLIP to model Ag nanoparticles on a graphene/Ni(111) support, achieving a 3–5 orders-of-magnitude speed-up over direct density-functional-theory calculations while retaining DFT-level accuracy for energies and forces \cite{rosen2025capturing}. After validating the model against DFT data and experimental adhesion energies, they showed that the idealised Platonic and Wulff geometries commonly assumed in computational catalysis are not appropriate for particles with effective diameters below about 8~nm. Instead, the supported nanoparticles adopt flattened Winterbottom-like shapes, with further global optimisation revealing even lower-symmetry structures. Another example is the study by Prasanth et al.~\cite{prasanth2023study}, which investigated the friction and wear behaviour of graphene-reinforced AA7075 nanocomposites using ML. By integrating experimental data with ML models, the researchers could predict how different material compositions and processing conditions would affect the composite's tribological properties. 
Surfaces and interfaces govern stability and function in nanomaterials by controlling surface energy, defect chemistry, functionalisation and morphology, and thus remain central to computational design of stable, synthesizable nanostructures \cite{mitchell2021nanoscale}. Device performance is likewise tied to substrate–film attributes, where atomic configurations at solids’ surfaces dictate electronic and chemical responses \cite{wan2024construction}. Yet accurate modelling of complex reconstructions (for example, silica) is still difficult. As shown in Figure~\ref{fig:fig1}a, empirical potentials mis-rank the Si(111) reconstructions: they favour the unreconstructed 1×1 surface and fail to identify the 7×7 dimer–adatom–stacking fault (DAS) structure as the lowest-energy configuration, even within the DAS family. The Si(111) surface illustrates both the structural richness and the modelling challenge: it exhibits (2n+1)×(2n+1) reconstructions, with the 7×7 structure featuring adatoms, 10-atom rings and stacking faults the accepted energetic ground state \cite{binnig19837}\cite{bartok2017machine}. Conventional empirical potentials can mis-rank these reconstructions. By contrast, an ML interatomic potential based on SOAP–GAP correctly identified Si(111)–7×7 as an energy minimum and reproduced the 19o Jahn–Teller dimer tilt on Si(100) in agreement with DFT Figure~\ref{fig:fig1}b \cite{bartok2018machine}, despite being trained on limited data an outcome that has made silicon surfaces a benchmark for many subsequent MLIP studies \cite{macisaac2024genetic} \cite{morrow2023validate}.

Nanoparticle stability is equally consequential for translation to applications \cite{jauffred2019plasmonic}. Gold nanoparticles (AuNPs) exemplify this: their use in optics, nanomedicine and catalysis hinges on shape- and temperature-dependent structural integrity \cite{daniel2004gold}. To resolve melting pathways at relevant sizes and timescales, Zeni \textit{et al}. built mapped-Gaussian-process MLIPs and ran molecular dynamics for AuNPs from 1–6 nm (147–6266 atoms) over an aggregate 2.4 ms well beyond electronic-structure limits \cite{zeni2021data}. An unsupervised analysis, combining a refined three-body local atomic-cluster-expansion descriptor (40-dimensional per atom) with hierarchical k-means, separated solid and liquid local environments and distinguished coordination-specific surface classes (Figure~\ref{fig:fig1}c). The simulations show surface-initiated melting that progresses inward (Figure~\ref{fig:fig1}d), consistent with experiment, and provide a tractable framework for interrogating nanoscale phase transitions and the role of surface structure in AuNP stability.

MLIPs have progressed from early neural-network fits for small molecules to high-dimensional neural-network potentials (HDNNPs) introduced by Behler and Parrinello \cite{behler2007generalized}, a DFT-trained framework that scales to systems with thousands of atoms.  This advance brings nanoscale systems such as nanoparticles, surfaces, defects and supported catalysts within reach at near first-principles fidelity and molecular-dynamics length and time scales. HDNNPs advance beyond earlier models by adding explicit long-range interactions. Most implementations include electrostatics, and some also account for dispersion, often via Grimme’s D3 correction \cite{grimme2010consistent}, as in TensorMol \cite{yao2018tensormol} and in other ML potentials such as PhysNet \cite{unke2019physnet}. In the HDNNPs, the electrostatic charges are not fixed by element or atom type; instead, they are predicted from each atom’s local chemical environment. This is a clear improvement over classical force fields, but it is still insufficient whenever long-range charge transfer occurs. In such cases, an atom’s charge can depend on structural or electronic changes far outside its local neighbourhood, and different global ionisation states can alter the overall charge distribution effects that purely local charge models cannot distinguish. 

\begin{figure}[H]
  \centering
  \includegraphics[width=\linewidth]{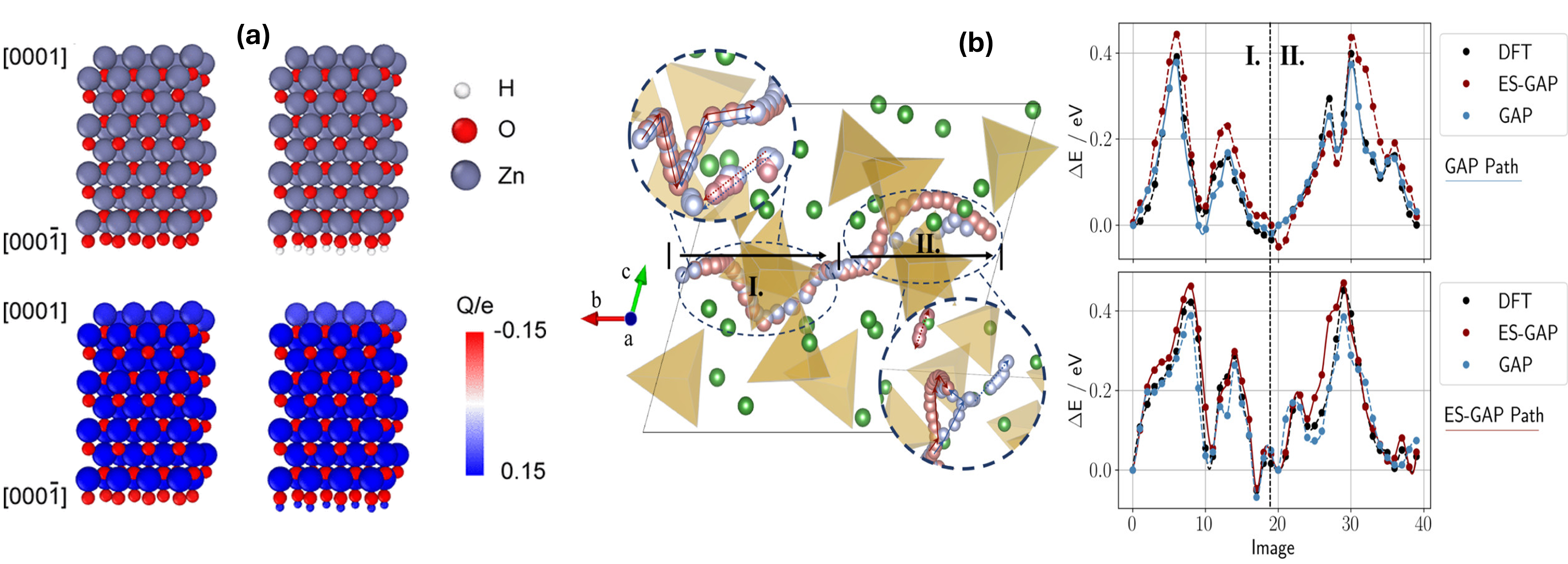}
  \caption{\textbf{Nonlocal charge transfer and ion-transport barriers captured by ML potentials.}
  \textbf{a}, Nonlocal charge transfer in a polar ZnO slab: (top) side view of a Zn-terminated face and the same slab with the O-terminated face hydrogenated; (bottom) corresponding atomic partial charges. Reproduced with permission from Ref.~\cite{behler2021four}.
  \textbf{b}, Lithium diffusion in Li$_7$P$_3$S$_{11}$: NEB minimum-energy paths for representative hops. DFT is the reference; GAP and ES-GAP closely reproduce the DFT energy profiles along the reaction coordinate. Barrier comparisons across symmetry-inequivalent hops show small, largely systematic deviations, with ES-GAP slightly improving agreement near saddle points. Reproduced with permission from Ref.~\cite{staacke2021role}.}
  \label{fig:fig2}
\end{figure}

For instance, Long-range charge transfer is common across chemistry, molecular biology, and materials science. Adding a neural network for electrostatics improves long-range behaviour, but the model still infers charges from local environments. Consequently, it cannot capture nonlocal charge transfer, which is common in interfacial systems. Figure~\ref{fig:fig2}a illustrates this on a polar ZnO slab: the top-layer Zn atoms have the same local geometry for both an oxygen-terminated surface and a hydrogen-terminated surface, yet the global dipole, electronic structure, and charge distribution differ \cite{behler2021four}. A strictly local charge model, therefore, assigns the same charges and atomic energy contributions to these Zn atoms in both terminations, which is qualitatively incorrect. The limitation is not unique to ML potentials; it also affects empirical force fields. A classic remedy is the charge equilibration method of Rappé and Goddard \cite{rappe1991charge}, which redistributes charges by solving equations based on atomic electronegativities and related quantities, and which underpins advanced force fields such as ReaxFF \cite{van2001reaxff}. In the ML domain, Ghasemi and co-workers introduced the charge-equilibration neural network (CNET), an MLIP that captures non-local electron transfer by coupling a neural network with an energy expression solved for charge equilibration \cite{ghasemi2015interatomic}.

To motivate when explicit long-range electrostatics are indispensable at the nanoscale, Staacke \textit{et al.} systematically compare short-range and electrostatics-augmented ML potentials for \ce{Li7P3S11} contrasting bulk transport with interfacial/defect-rich conditions \cite{staacke2021role}. Staacke \textit{et al.} probe when long-range electrostatics must be made explicit in atomistic ML potentials by training two GAP models for the solid-state electrolyte \ce{Li7P3S11}: a strictly short-range GAP and a hybrid ES-GAP with a fixed-charge electrostatic baseline. In homogeneous bulk, both models reproduce lithium migration barriers (NEB) and MD conductivities with only modest, systematic offsets, consistent with electrostatic contributions averaging out in an isotropic crystal. As illustrated in Figure~\ref{fig:fig2}b, both a short-range GAP and an electrostatics-augmented ES-GAP reproduce the \emph{nanoscale} lithium migration landscape in bulk \ce{Li7P3S11}: minimum-energy paths and barriers from NEB closely track DFT, with only modest, systematic offsets. This supports the view that in homogeneous bulk the long-range contributions largely average out, whereas interfacial or field-driven cases require explicit electrostatics. Once nanoscale symmetry is broken by applied fields or Frenkel defects that mimic electrode|electrolyte interphases the differences become qualitative: only the electrode-electrolyte interphases the differences become qualitative: only the electrostatics-aware model captures the field-dependent stabilisation or destabilisation interfacial fields. Methodologically, a simple fixed-charge baseline reduces long-range force nonlocality but can mis-handle chemically distinct sulfur sites, suggesting that floating/learned charges or explicit charge-equilibration would be preferable for interfacial nanoscale physics. 

Furthermore, materials informatics is making strides in understanding chemical processes at the nanoscale~\cite{wang2023machine}. Wang \textit{et al}. conducted a study on the sulfur poisoning mechanism of LSCF cathode materials in the presence of \ce{SO2}. By combining computational and experimental methods with ML, the researchers could unravel the intricate interactions at the atomic level that lead to material degradation. This understanding is crucial for developing more durable materials for energy applications, such as fuel cells and batteries, where chemical stability is a significant concern. Similarly, for energy materials, Wang \textit{et al.}~\cite{wang2022machine} utilised ML to guide dopant selection for metal oxide-based photoelectrochemical water splitting, focusing on \ce{Fe2O3} and CuO. The study demonstrated how ML could accelerate the identification of optimal dopant combinations, enhancing the efficiency of photoelectrochemical processes, which expands the scope of exploration beyond what is feasible through traditional experimental methods alone.

Moreover, materials informatics is playing a crucial role in the development of environmentally friendly materials and processes. Wang et al.~\cite{wang2023machine} applied ML to predict and optimise the performance of thin-film nanocomposite membranes used in organic solvent nanofiltration. By accurately predicting membrane performance based on material composition and structure, ML models can guide the design of more efficient and selective membranes, contributing to the development of greener industrial processes.

The success of materials informatics relies heavily on the synergy between experimental data and computational modelling. Experimental data provides the foundation for training ML models, while computational models help to simulate and predict material behaviour, guiding further experimental efforts.

\subsubsection{Open Challenges }
MLIPs have proved useful for modelling nanoscale systems; however, challenges in data coverage, transferability, long-range electrostatics and charge transfer, robustness under extrapolation, and reproducibility must be addressed before they can be used routinely for interfacial and other complex systems. 

For instance, model interpretability is a significant challenge in materials informatics~\cite{oviedo2022interpretable},~\cite{butler2021interpretable}. Complex models, such as deep neural networks, often act as "black boxes," making it difficult to understand how predictions are made, which can be a barrier to the adoption of ML models in materials science, where understanding the underlying physical mechanisms is crucial. Li \textit{et al}.~\cite{li2024forecasting} discussed this issue in their work on forecasting the strength of nanocomposite concrete, noting that while interpretable models like decision trees are more transparent, they may not capture the full complexity of the material behaviours as effectively as more complex models.

Another challenge is the integration of experimental and computational data. Wang \textit{et al}.~\cite{wang2023sulfur}, in their study on the sulfur poisoning mechanism of LSCF cathode materials, highlighted the difficulties in combining experimental data with computational models to create a comprehensive ML framework. While ML models can be trained on either type of data, integrating both into a single predictive model remains challenging due to differences in data formats, scales, and underlying assumptions which limits the ability to fully leverage all available data for more accurate and robust predictions.

Scalability is another unresolved problem, particularly in applications requiring large-scale simulations or high-throughput experimentation. Choi \textit{et al}.~\cite{Choi2024} addressed this in their work on accelerating phase-field simulations of microstructure evolution in additive manufacturing. They noted that while ML can significantly reduce computational costs, scaling these methods to handle the vast design spaces of advanced materials remains a challenge. This issue is compounded by the need for real-time predictions and optimisations in manufacturing processes, where delays can be costly.

Much of the current research is based on computationally generated data, such as DFT calculations, which may not always accurately reflect real-world conditions. For example, the application of ML to nanoporous materials is complicated by the need for simultaneous optimisation of the solid phase structure and the nanospace, a task that requires extensive experimental data. The authors emphasise that more experimental data are needed to develop "synthesizable" materials, and the current dependence on computational data poses limitations on the generalizability of models~\cite{anand2023exploiting}.

Uncertainty quantification is also a crucial area that needs more attention. While some studies, like that of Wang \textit{et al}.~\cite{wang2022machine} on dopant selection in photoelectrochemical water splitting, have begun to incorporate uncertainty quantification into their models, this practice is not yet widespread. Many ML models provide point predictions without accounting for the uncertainty inherent in the data or the model itself. This limitation can lead to overconfidence in predictions and potentially costly errors in material design and application.
Transferability of ML models across different material systems remains a significant challenge. For instance, Jin \textit{et al}.~\cite{jin2023machine} noted that models developed for nickel–graphene nanocomposites may not be directly applicable to other types of nanocomposites without significant retraining and adaptation, which however, limits the broader applicability of ML models, requiring extensive data collection and model development for each new material system.

\subsection{Mesoscale learning}
Mesoscale (10 µm–1 mm) materials bridge micro- and macro-scale material properties, playing a crucial role in materials design and optimisation \cite{fish2021mesoscopic}. The field has emerged as a critical direction in materials informatics, driven by the high computational cost and structural complexity, including phase-field, coarse-grained, and microstructure-resolved simulations. ML offers new opportunities at this scale by enabling data-driven surrogates, latent representations, and operator learning frameworks that capture mesoscale evolution directly from high-fidelity simulations, circumventing the need for repeated numerical solution of governing equations. At the mesoscale, materials exhibit unique behaviours that significantly influence their macroscopic properties, such as mechanical strength, thermal conductivity, and elasticity. Understanding and manipulating these behaviours are essential to create innovative materials for various industries, including aerospace, automotive, and biomedical engineering~\cite{olfatbakhsh2022highly}. The most prominent application of mesoscale materials informatics is in the design of metamaterials, engineered materials with properties that do not naturally occur. These materials are structured at the mesoscale to achieve specific functionalities, such as negative refractive indices or enhanced mechanical performance. Research in this area, such as the study by Kulagin et al.~\cite{kulagin2023lattice}, demonstrates how Bayesian optimisation can be employed to design lattice metamaterials with mesoscale motifs, significantly improving their elastic properties by manipulating the geometry of unit cells.

One notable area of application is the development of metamaterials, engineered materials with unique properties that are not found in nature. The design of these materials requires a deep understanding of the interactions between microstructural features and macroscopic properties, which is where data-driven approaches excel. For example, Kulagin \textit{et al}.~\cite{kulagin2023lattice} used Bayesian optimisation to enhance the elastic properties of lattice metamaterials by manipulating the geometry of mesoscale motifs, demonstrating how data-driven models can efficiently explore vast design spaces to identify optimal configurations. Similarly, in the realm of fracture modelling, Panda et al.~\cite{panda2020mesoscale} demonstrated the use of ML to emulate fine-scale physics, particularly in the context of fracture propagation in brittle materials. Their study developed a framework that uses ML to predict continuum-scale parameters by emulating complex mesoscale phenomena, such as the evolution of fractures within a material. This approach significantly reduces computational costs compared to traditional high-fidelity models while maintaining accuracy, showcasing the potential of ML in bridging the gap between mesoscale modelling and macroscopic predictions. The adoption of material informatics approaches on the mesoscale has brought numerous benefits, particularly in terms of efficiency, accuracy, and the ability to handle complex, high-dimensional data. 

Other critical areas of research involve multi-scale modelling, which integrates mesoscale information with atomistic simulations to predict macroscopic material behavior, which is particularly useful in optimising the mechanical properties of materials, as demonstrated in studies that combine DFT with mesoscale modelling~\cite{hasan2022data}. By incorporating mesoscale features such as grain boundaries and defects into the models, researchers can accurately predict the material’s response to various mechanical loads, enabling the design of materials with superior performance. ML and data-driven methods have also become integral to mesoscale materials informatics, enabling efficient exploration of vast design spaces and the identification of optimal material configurations. For example, ML algorithms have been used to optimise the elastic properties of cubic microstructures by analysing large datasets, identifying patterns that lead to desirable mechanical characteristics, and reducing the need for extensive physical experimentation~\cite{hasan2022data}. These techniques are crucial for accelerating the materials discovery process, particularly in cases where traditional trial-and-error methods would be prohibitively time-consuming.

A representative example of learning on the mesoscale is the use of ML to accelerate phase-field simulations\cite{zhao2023understanding} of microstructure evolution, which are widely used mesoscale models for phenomena such as spinodal decomposition and grain growth. In this approach, high-fidelity phase-field simulations are first performed to generate time-resolved microstructure fields (e.g., spatial distributions of phase or composition). These simulation outputs, which are computationally expensive to obtain, are then used as training data for ML surrogate models. For instance, Montes de Oca Zapiain \textit{et al.} \cite{montes2021accelerating} trained data-driven models to learn the temporal evolution of microstructures directly from phase-field data, enabling rapid prediction of future microstructure states without explicitly solving the governing partial differential equations at each time step. The trained surrogate reproduced key morphological features such as domain size and phase connectivity while achieving orders-of-magnitude reductions in computational cost. 

\begin{figure}[H]
  \centering
  \includegraphics[width=\linewidth]{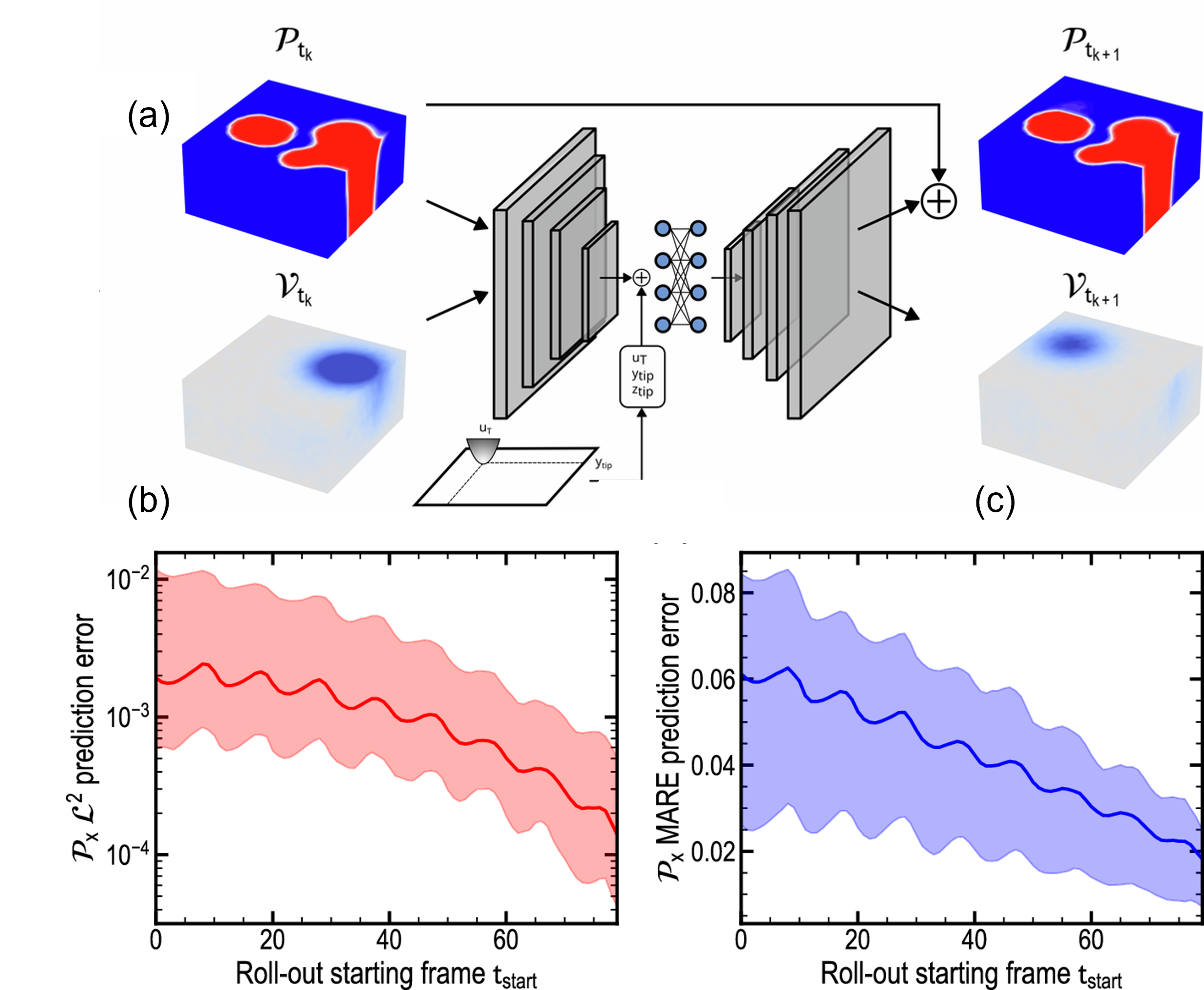}
  \caption{ML surrogate architecture and long-horizon rollout accuracy for tip-induced ferroelectric switching. (a) Schematic of the surrogate time-stepping framework: the microstructural state at time $t_k$, comprising the polarization field $\mathbf{P}_{t_k}$ and electrostatic potential $V_{t_k}$, is combined with electrically imposed boundary-condition descriptors (tip position $(y_{\mathrm{tip}}, z_{\mathrm{tip}})$ and applied voltage $u_T$) to predict the subsequent state at $t_{k+1}$. (b) Evolution of the $\mathcal{L}^2$ prediction error for the polarization field during unrolled simulations, shown as a function of the rollout starting frame $t_{\mathrm{start}}$. (c) Corresponding macro-average relative error (MARE) for the polarization field under the same rollout conditions. In (b,c), statistics are computed over 200 independent test trajectories; solid lines denote the mean error, and shaded regions indicate the interquartile range (25th--75th percentiles). Reproduced with permission from Ref.~\cite{alhada2024machine}.}
  \label{fig:pf_surrogate}
\end{figure}

Alhada-Lahbabi \textit{et al.} \cite{alhada2024machine} showed that a surrogate can make mesoscale 3D phase-field simulations of ferroelectric domain evolution in PbZr$_x$Ti$_{1-x}$O$_3$ (PZT) thin films usable in inverse problems by learning an explicit time-stepping map for the polarisation and electrostatic potential fields. Rather than emulating a single fixed protocol, the surrogate is conditioned on time-dependent electrical boundary conditions that represent the probe action (tip position, applied bias, and application time), so the same model can be rolled out across different switching sequences Figure~\ref{fig:pf_surrogate}a. This "state + boundary-condition controls next state" formulation is the key step that connects phase-field mesoscale physics to optimization or control loops, where thousands of forward evaluations are typically required. 
The authors report stable long-horizon rollouts with macro-average relative error below $\sim$10\% and speedups up to $\sim$2500$\times$ compared with the direct solver Figure~\ref{fig:pf_surrogate}b and c, which is the regime needed for practical design iterations rather than one-off simulations. The approach also highlights a broader lesson for mesoscale materials informatics: accuracy should be assessed on iterated trajectories (error accumulation and stability), not only on one-step prediction. Conceptually, it complements earlier phase-field acceleration studies for microstructure evolution that generally assume fixed loading paths \cite{montes2021accelerating} \cite{peivaste2022machine}, and it sits alongside operator-learning approaches that aim to learn evolution operators for mesoscale fields across wider parametric spaces \cite{lu2021learning} \cite{oommen2024rethinking}.
\begin{figure}[!t]
\end{figure}

Beyond direct surrogate replacement of mesoscale solvers, recent studies have focused on learning low-dimensional representations of mesoscale dynamics to predict long-time microstructure evolution. In these approaches, high-dimensional microstructure fields generated by phase-field simulations are first compressed into a latent space using autoencoders, capturing essential mesoscale features such as phase topology and characteristic length scales. The temporal evolution in this latent space is then learned using recurrent neural networks or neural operators. For example, Hu \textit{et al.} \cite{hu2025accelerating} combined autoencoders with recurrent networks to predict the evolution of microstructures during Ostwald ripening, achieving accurate long-time predictions at a fraction of the computational cost of conventional phase-field simulations. Similarly, operator-learning frameworks have been applied to learn the functional mapping between microstructure states at different times, enabling generalisation across initial conditions. 

In addition to computational modelling and ML, experimental methods such as high-throughput screening and in-situ characterisation are also vital components of mesoscale materials informatics \cite{shahzad2024accelerating}. These techniques allow for the rapid evaluation of material properties and the identification of promising material candidates. For instance, in the field of nanoporous materials, databases have been constructed to facilitate the application of materials informatics in predicting properties and optimising synthesis conditions~\cite{stewart2022recent}. The integration of computational, experimental, and data-driven approaches in mesoscale materials informatics has led to significant advancements in the field. Researchers are now able to design materials with tailored properties for specific applications, from lightweight structural components to high-performance catalysts. As the field continues to evolve, the development of more sophisticated models and algorithms will further enhance our ability to predict and optimise material behaviour at the mesoscale, paving the way for the next generation of advanced materials~\cite{xiao2020machine}.

Furthermore, the ability to handle large datasets and extract meaningful insights from them is a cornerstone of materials informatics. This capability is particularly important in mesoscale studies, where the interactions between microstructural features and material properties are complex and multidimensional. For instance, the work by Agarwal \textit{et al}.~\cite{agarwal2024multiscale} on predicting the mechanical properties of woven fabric composites showcases how ML models can be trained on microstructural data to predict macroscopic properties with high accuracy, which improves the predictive capabilities of models but also enhances the interpretability of the results, making it easier to understand the underlying mechanisms that drive material behaviour.

From a broader perspective, the difficulty of learning at the mesoscale can be understood through the lens of mesoscience, which emphasises that mesoscale structure often emerges from the interaction and competition of multiple physical mechanisms rather than a single dominant process. Guo \textit{et al.}~\cite{guo2019complexity} argue that many deep-learning models struggle to reveal underlying physical logic because they treat complex systems as purely statistical input--output mappings, without explicitly accounting for changes in dominant mechanisms across operating conditions. In mesoscale materials systems, coupled processes such as transport, deformation, phase transformation, or electrostatics can alternately dominate or interact, leading to heterogeneous spatiotemporal patterns and abrupt changes in behaviour~\cite{huang2016mesoscale}, which may provide a conceptual rationale for recent hybrid strategies that combine physics-aware decomposition with data-driven learning, particularly in cases where end-to-end models struggle with stability, interpretability, or extrapolation across processing conditions.

\subsubsection{Open Challenges }
Despite significant advancements in mesoscale materials informatics, several challenges and limitations remain that current state-of-the-art approaches have not fully addressed. 

One of the primary challenges in mesoscale materials modelling is the integration of models across multiple scales. While multi-scale modelling is essential for understanding the behaviour of materials from the atomic to the macroscopic level, the computational costs associated with incorporating fine-scale physics into higher-scale models are often prohibitive. Panda et al. [70] discuss the challenges in scaling mesoscale information to the continuum or macroscale, particularly when considering uncertainty.
    
Computational costs and scalability remain significant challenges in mesoscale materials informatics. Full-field simulations of mesoscale phenomena are often computationally demanding, particularly for transient and multiphysics problems involving coupled processes such as diffusion, phase transformation, and deformation. Although high-performance computing (HPC) has significantly expanded accessible problem sizes, it remains insufficient for applications that require repeated evaluations, long-time evolution, or extensive exploration of parameter spaces. Martirossyan \textit{et al.}~\cite{martirossyan2024snapshot} highlight the difficulty of resolving spatio-temporal coupling across multiple transient timescales, which is common in mesoscale diffusion--mechanics problems. As a result, reduced-order models (ROMs) and surrogate approaches are frequently employed to improve tractability; however, these methods may compromise accuracy when applied to strongly nonlinear regimes or conditions outside their training domain.

Challenges in emulating fine-scale physics\textbf{.} Building accurate emulators for fine-scale physics remains challenging when mesoscale models must resolve microstructural heterogeneity and provide reliable uncertainty quantification. This issue is particularly evident in mesoscale fracture modelling, where damage evolution and failure are strongly influenced by local microstructural features. Panda \textit{et al.} \cite{panda2020mesoscale} demonstrate that, although ML surrogates can approximate fracture responses, constructing well-calibrated emulators for time-dependent quantities such as total damage and maximum stress remains difficult, highlighting a key limitation for predictive mesoscale modelling. 

Limitations in Current ML Models. Although ML models have proven to be powerful tools in material informatics, they are not without limitations. For instance, current ML models often struggle with handling the high dimensionality and complexity of materials data, particularly when multiple interacting variables are involved. Additionally, there is an ongoing challenge in selecting the "right" type of neural network for different types of physical problems. The need for optimal sampling strategies to reduce computational costs and avoid generating unnecessary data samples is another area that requires further research. The lack of universal frameworks for integrating physical laws into ML models, which could improve their interpretability and reliability, is also a significant limitation~\cite{stewart2022recent}.

While mesoscale materials informatics has made significant strides in recent years, numerous challenges remain to be addressed to fully realise its potential. These challenges include the integration of multi-scale models, the need for high-quality experimental data, the computational costs associated with simulating complex phenomena, the difficulty of emulating fine-scale physics, and the limitations of current ML models. Addressing these issues will require continued innovation in both computational methods and experimental techniques, as well as the development of more robust and scalable models.

\subsection{Microscale learning}
At the microscale, machine learning has enabled automated and quantitative analysis of experimentally resolved microstructural features obtained directly from microscopy. Deep convolutional neural networks have been widely applied to pixel-wise segmentation of complex phase microstructures, often outperforming classical thresholding and morphology-based approaches for identifying grains and phases in scanning electron microscopy (SEM) and scanning transmission electron microscopy (STEM) images. Representative examples include three-dimensional segmentation of polycrystalline ceramics from serial sectioning data by Hirabayashi \textit{et al.} \cite{hirabayashi2024deep} and few-shot segmentation frameworks for STEM images developed by Akers \textit{et al.} \cite{akers2021rapid}. Neural approaches have also been used to classify steel microstructures directly from raw micrographs, providing more objective and reproducible alternatives to manual expert annotation, as demonstrated by Azimi \textit{et al.} \cite{azimi2018advanced}, and to segment complex phase distributions in structural steels using fully convolutional models. Beyond segmentation and classification, microstructure-to-property mapping at the microscale has advanced through graph neural networks trained on experimentally derived microstructural graphs, enabling improved prediction of fatigue damage accumulation relative to traditional phenomenological models \cite{thomas2023materials}. More recently, vision-transformer-based representations have been explored to learn task-agnostic microstructure descriptors that can be transferred across multiple property-prediction tasks \cite{whitman2025machine}. Together, these studies demonstrate how machine learning can extract and exploit structure–property relationships directly at the microscale, where explicit microstructural features such as grain morphology, phase topology, and local connectivity govern material response. 

At the microscale, ML methods have been used primarily to analyse experimentally resolved microstructures obtained from optical and electron microscopy. Convolutional neural networks have enabled pixel-wise segmentation of complex multiphase microstructures that cannot be reliably separated using grayscale thresholding or hand-crafted image features. A representative example is the work of Durmaz \textit{et al.} \cite{durmaz2021deep}, who addressed the segmentation of lath-shaped bainite in complex-phase steels by training U-Net architectures on a limited number of light-optical and scanning-electron micrographs. To establish objective ground truth, correlative electron backscatter diffraction (EBSD) data were used during annotation, allowing reliable discrimination of bainite and ferrite based on crystallographic orientation and intragranular misorientation metrics \cite{li2018quantification} \cite{chen2018phase}. The resulting models achieved segmentation accuracies comparable to expert assessments, despite being trained on fewer than fifty images, demonstrating that carefully curated microscale datasets can suffice when image variance is controlled. Related studies have shown that similar deep-learning frameworks can outperform classical computer-vision pipelines for metallographic segmentation and classification tasks \cite{decost2019high}\cite{bulgarevich2018pattern}, while also enabling quantitative extraction of phase fractions and morphological descriptors relevant to mechanical performance. Importantly, the interpretability of the learned representations was examined using gradient-based and filter-level visualisation techniques, revealing that the networks rely on physically meaningful features such as grain boundaries, carbide morphology, and phase topology \cite{selvaraju2017grad}. ML-assisted microstructure establishes inference as a practical tool at the microscale, where explicit microstructural features are directly observable and strongly linked to material behaviour. 

Furthermore, Wijaya \textit{et al}.~\cite{wijaya2024analyzing} combined multiple ML methods to analyse microstructure relationships in porous copper. The study integrated experimental data from various characterisation techniques with ML models to uncover the links between microstructural features and material performance. This multi-method approach is particularly powerful in materials informatics, as it allows for a comprehensive understanding of material behaviour across different length scales. Another active area at the microscale is the integration of simulations from atomistic and continuum paradigms. Recently, researchers integrated atomistic simulations with continuum models to predict the behaviour of complex materials across different length scales, providing a more comprehensive understanding of material behaviour, improving the accuracy of predictions and aiding in the design of materials with specific properties.

\subsubsection{Open Challenges }
The challenges described are intrinsically tied to the microscale, as defined by the use of spatially resolved microstructural data obtained from microscopy and diffraction techniques. Data scarcity arises because microscale characterization methods such as SEM, TEM, EBSD, and 3D FIB-SEM tomography are experimentally intensive and slow, a limitation explicitly identified in foundational microstructure informatics work \cite{kalidindi2015materials}\cite{bostanabad2018computational}. The problem of ground-truth definition is microscale-specific because entities like grains, phase boundaries, and defects do not have unique physical definitions independent of segmentation rules, image contrast, or operator decisions, an issue repeatedly highlighted in microstructure segmentation and EBSD analysis studies \cite{decost2015computer}\cite{holm2020overview}.

Similarly, domain shift is a direct consequence of microscale imaging physics: variations in microscope type, accelerating voltage, detector geometry, polishing, and etching change image appearance without altering the underlying material, causing ML models to fail across datasets a problem explicitly discussed in microscopy-based ML studies.\cite{ziletti2018insightful}. The difficulty of defining suitable representations stems from the fact that microstructures are heterogeneous, multiscale, and topological, where spatial correlations, connectivity, and orientation relationships control behaviour, distinguishing microscale informatics from composition- or property-only learning \cite{bostanabad2018computational}\cite{li2018transfer}.

Noise, limited fields of view, and reconstruction artefacts are inherent to microscale measurements and strongly affect inferred statistics such as grain size distributions or phase connectivity, particularly in 3D datasets, as emphasised in computational microstructure reconstruction literature\cite{chen2016stochastic}. Class imbalance reflects the physical rarity of microscale phenomena such as crack initiation sites or critical inclusions, which has been noted as a key obstacle in defect detection and microstructural classification tasks.\cite{moreh2024deep}. The difficulty in linking microscale descriptors to macroscopic properties is a direct consequence of attempting to bridge microstructural length scales to continuum behaviour, an ill-posed problem widely discussed in structure–property and inverse microstructure design studies.\cite{agrawal2016perspective}.

\subsection{Methods and Toolchains }
Across the different length scales considered in this Review, a broadly similar family of ML methods is used; what changes with scale is the structure of the data, the representation and the physical questions being asked. Artificial neural networks (ANNs) are widely employed to capture nonlinear relationships between processing parameters, structure and properties; for example, Youshia \textit{et al}. used ANNs to predict the particle size of polymeric nanoparticles from processing conditions, enabling accurate, data-driven guidance for experimental efforts~\cite{youshia2017artificial}. Support vector machines (SVMs) are a common choice for regression and classification tasks when datasets are of modest size: in the work of Huang \textit{et al} on carbon nanotube-reinforced cement composites, SVMs were used to assess mechanical properties, successfully identifying patterns that correlate with strength and thereby supporting materials optimisation~\cite{huang2021data}. Tree-based models, including decision trees and random forests, remain popular because of their robustness and interpretability. Li \textit{et al}. employed decision trees to forecast the strength of nanocomposite concrete containing carbon nanotubes, using the graphical structure of the model to gain insight into which microstructural and compositional features most strongly influence the response~\cite{li2024forecasting}. Gaussian processes (GPs) are favoured in applications where uncertainty quantification is important. For example, Wang \textit{et al}. used GP-based models to guide dopant selection in metal-oxide photoelectrochemical water-splitting systems, exploiting the associated confidence intervals to steer experiments efficiently in composition space~\cite{wang2022machine}. For high-dimensional or unstructured inputs such as microstructure images, convolutional neural networks (CNNs) and related deep architectures are increasingly used: Wijaya \textit{et al}. combined CNNs with other ML methods to analyse microstructure–property relationships in porous copper, extracting informative features from images to gain deeper insight into performance~\cite{wijaya2024analyzing}. Forni \textit{et al}. employed convolutional neural networks on image-based representations to predict electronic and physical properties of nanographenes~\cite{Forni2024}. Graph neural networks (GNNs) and other message-passing models offer a complementary route for materials with complex, hierarchical structures, where they can capture spatial correlations and topology in ways that are difficult for hand-crafted descriptors~\cite{stewart2022recent}. Reinforcement-learning (RL) approaches are beginning to appear in process optimisation and microstructure control; Choi \textit{et al}. integrated ML predictions with phase-field simulations to optimise microstructure evolution in additive manufacturing, demonstrating how RL can learn strategies for selecting processing parameters in real time~\cite{choi2024accelerating}.

The overall workflow in a typical materials-informatics study is similar across nanoscale, mesoscale and microscale problems. Data are assembled from experiments, simulations or existing databases, followed by feature engineering, model training and validation. High-throughput experiments and simulations generate large datasets that are processed using a range of supervised-learning methods such as regression and classification to predict properties from structural or process descriptors, for example, grain size, phase distribution or microstructure statistics in mesoscale studies of mechanical strength and thermal conductivity~\cite{selvaraj2023multiscale}. Unsupervised techniques, including clustering and principal component analysis (PCA), are used to discover latent patterns in such datasets, for instance, to categorise microstructural images into classes that correlate with distinct properties or deformation modes~\cite{martirossyan2024snapshot}. Active-learning strategies build on probabilistic models, such as GPs, to select the most informative experiments or simulations, thereby reducing the number of data points needed to reach a given level of accuracy~\cite{panda2020mesoscale,choudhary2020joint}. RL and other sequential-design methods are used when the goal is to optimise a process or trajectory rather than a single configuration, as in the additive-manufacturing example above~\cite{choi2024accelerating}.

Implementation of these methods is enabled by a set of general-purpose machine-learning libraries and domain-specific software toolchains. Packages such as \textsc{scikit-learn}, TensorFlow and Keras provide the core optimisation algorithms and model abstractions used for many supervised and unsupervised methods, and are routinely employed to prototype and deploy ANN, SVM, tree-based and GP models in materials applications. For domain-specific tasks, libraries such as \textsc{matminer} and the Materials Project interface provide specialised tools for materials data mining and analysis, including feature extraction from crystal structures and computed properties, which makes it easier to integrate materials-science knowledge into ML models. At the atomistic level, molecular-dynamics engines such as LAMMPS and HOOMD-blue are widely used to simulate the behaviour of materials, and are often coupled to ML frameworks like PyTorch or TensorFlow when training interatomic potentials or surrogate models. Frameworks such as JARVIS-ML integrate various ML models with DFT and MD simulations to automate parts of the materials-discovery workflow, supporting predictive models that span electronic, structural and mechanical properties~\cite{panda2020mesoscale,choudhary2020joint}. Quantum-chemical and mesoscale calculations are frequently carried out using tools such as Gaussian and Materials Studio, which can be combined with ML models to explore large design spaces more efficiently, for example, by using finite element analysis (FEA) plus ML to predict composite mechanical properties from microstructure, or by coupling MD and ML to model gas transport through nanoporous materials.

Underpinning these workflows is a growing ecosystem of materials databases and data platforms that provide curated structures and properties. The Materials Project (MP) is a widely used resource that offers computed crystal structures, phase diagrams and a wide range of electronic, structural and thermodynamic properties, together with high-throughput, automated workflows built on \textsc{pymatgen}, \textsc{custodian}, FireWorks and related tools for composition and structure design, crystal- and electronic-structure analysis, thermodynamic and kinetic interpretation, and microstructure characterisation. AFLOW is another large-scale materials-discovery database, containing millions of calculated materials and hundreds of millions of associated properties, and is heavily used for screening candidate compounds~\cite{calderon2015aflow}. The Materials Genome Initiative (MGI) has framed many of these efforts, emphasising the integration of multiscale, high-throughput computation, experiment and data to reduce the time and cost of materials development~\cite{agrawal2016perspective}. Knowledge-enabled systems such as MATCALO combine ML with machine-interpretable semantic knowledge to represent relationships between materials, processes and properties and to support informed exploration of new materials spaces~\cite{picklum2019matcalo}. 
Many open-source tools, such as ASE (Atomic Simulation Environment), PySCeS (Python Simulator for Cellular Systems), and Matplotlib, are widely used for materials modelling, simulation, and data analysis. The development and use of these tools promote transparency and reproducibility in research. 
Crystallographic resources, including the Bilbao Crystallographic Server, the Cambridge Structural Database (CSD), the Crystallography Open Database (COD) and the Inorganic Crystal Structure Database (ICSD), provide symmetry information, Wyckoff positions and curated crystal structures for inorganic, organic and organometallic compounds, and are used extensively to generate and validate structural inputs for ML models. Platforms such as Materials Cloud, Materials Atlas and the Materials Platform for Data Science (MPDS) add further capabilities for hosting, visualising and querying computational and experimental results, including phase diagrams and physical properties extracted from the literature. NOMAD offers a repository and analysis platform for atomistic simulations and multiscale modelling data, with an emphasis on quantum-chemical properties and long-term preservation~\cite{scheidgen2023nomad}. Finally, workflow engines such as AiiDA provide an automated, provenance-aware infrastructure for computational science, with advanced error handling and restart capabilities that are particularly valuable in high-throughput and ML-driven studies. Together, these methods, software tools and data resources form a shared toolkit that can be adapted to atomistic, nano-, meso-, and micro-to-continuum problems; which elements are chosen in practice depends less on the length scale than on data availability, noise, input modality and the need for interpretability or uncertainty quantification.

\begin{table}[htbp]
\centering
\caption{Summary of tools and software used across different length scales in materials informatics}
\begin{tabular}{|l|c|c|c|p{4.5cm}|}
\hline
\textbf{Tool/Software} & \textbf{Nanoscale} & \textbf{Mesoscale} & \textbf{Microscale} & \textbf{References} \\ \hline
Artificial Neural Networks (ANNs) & ✓ &  &  & \cite{youshia2017artificial}\\ \hline
Support Vector Machines (SVMs) & ✓ &  &  & \cite{huang2021data} \\ \hline
Decision Trees / Random Forests & ✓ &  &  & \cite{li2024forecasting} \\ \hline
Gaussian Processes (GPs) & ✓ &  &  & \cite{wang2022machine} \\ \hline
Convolutional Neural Networks (CNNs) & ✓ & ✓ &  & \cite{wijaya2024analyzing,stewart2022recent} \\ \hline
Reinforcement Learning (RL) & ✓ & ✓ &  & \cite{choi2024accelerating,panda2020mesoscale} \\ \hline
Active Learning &  & ✓ &  & \cite{panda2020mesoscale,choudhary2020joint} \\ \hline
Principal Component Analysis (PCA) &  & ✓ &  & \cite{martirossyan2024snapshot} \\ \hline
Clustering &  & ✓ &  & \cite{martirossyan2024snapshot} \\ \hline
Graph Neural Networks (GNNs) &  & ✓ &  & \cite{stewart2022recent} \\ \hline
Finite Element Analysis (FEA) &  & ✓ &  & \cite{selvaraj2023multiscale} \\ \hline
Scikit-learn & ✓ &  &  & \cite{pedregosa2011scikit}\\ \hline
TensorFlow & ✓ & ✓ &  & \cite{abadi2016tensorflow}\\ \hline
Keras & ✓ &  &  & \cite{chollet2015keras}\\ \hline
PyTorch &  & ✓ &  & \cite{paszke2019pytorch}\\ \hline
Matminer & ✓ &  &  & \cite{ward2018matminer}\\ \hline
LAMMPS &  & ✓ &  & \cite{plimpton1995fast}\\ \hline
HOOMD-blue &  & ✓ &  & \cite{anderson2020hoomd}\\ \hline
JARVIS-ML &  & ✓ &  & \cite{choudhary2020joint} \\ \hline
Gaussian (software) &  & ✓ &  & \cite{frisch2016gaussian16}\\ \hline
Materials Studio &  & ✓ &  & \cite{materials_studio}\\ \hline
Materials Project (MP) &  &  & ✓ & \cite{jain2013commentary}\\ \hline
MATCALO &  &  & ✓ & \cite{picklum2019matcalo} \\ \hline
Automatic FLOW (AFLOW) &  &  & ✓ & \cite{calderon2015aflow} \\ \hline
Bilbao Crystallographic Server &  &  & ✓ & \cite{aroyo2006bilbao}\\ \hline
Cambridge Structural Database (CSD) &  &  & ✓ & \cite{groom2016cambridge}\\ \hline
Crystallography Open Database (COD) &  &  & ✓ & \cite{gravzulis2012crystallography}\\ \hline
JARVIS &  &  & ✓ & \cite{choudhary2020joint}\\ \hline
Materials Cloud &  &  & ✓ & \cite{talirz2020materialscloud}\\ \hline
Materials Platform for Data Science (MPDS) &  &  & ✓ & \cite{mpds}\\ \hline
Inorganic Crystal Structure Database (ICSD) &  &  & ✓ & \cite{hellenbrandt2004inorganic}\\ \hline
Novel Materials Discovery (NoMAD) &  &  & ✓ & \cite{scheidgen2023nomad} \\ \hline
AiiDA &  &  & ✓ & \cite{pizzi2016aiida}\\ \hline
\end{tabular}
\label{tab:tools}
\end{table}

\FloatBarrier

\section{Bridging the scales: Common language and standards}
\begin{figure}[H]
    \centering
    \includegraphics[width=0.8\linewidth]{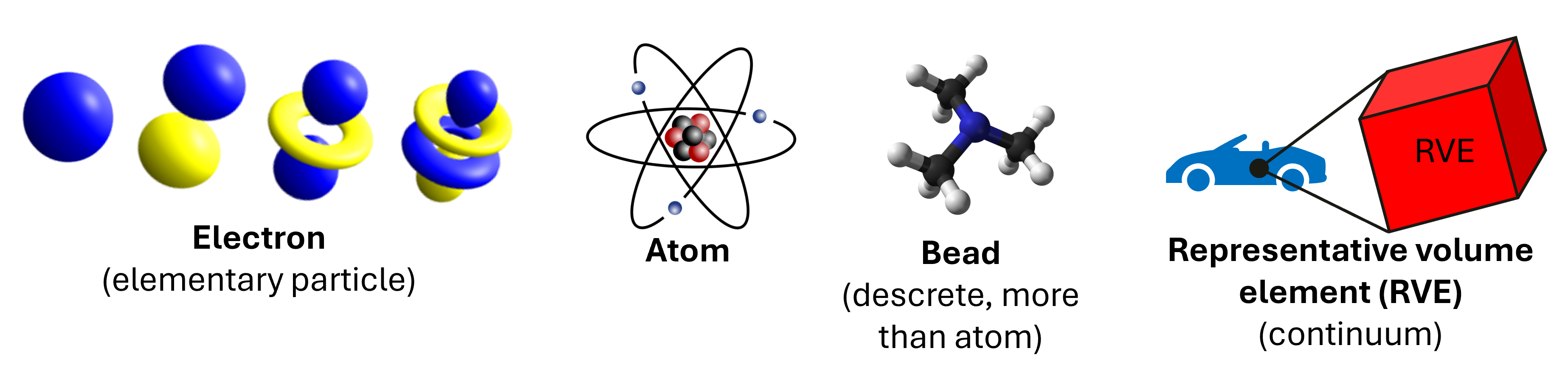}
    \caption{Materials model entities. The materials model entities are electrons and atoms at nanoscale, beads (discrete unit consisting of more than one atom) at mesoscale and continuum volume at the at microscale. }
    \label{fig:materials-entities}
\end{figure}

The previous three sections focus on the use of ML and AI at three different length scales. Materials models at these three scales predict the behaviour of distinct materials entities, which are self-contained, internally frozen, structureless representational units of the material shown in Figure~\ref{fig:materials-entities}. 

The materials entities at the nanoscale are the electron (and other spin-½ fermions) and the atom \cite{solov2022dynamics}. Electrons are fully quantum mechanical in nature, and their wave-nature is described by the Dirac equation, or the Schrödinger equation in the common non-relativistic case. The dynamics of atoms, on the other hand, can be described with Newton’s equations of motion, where the interaction between atoms is expressed with interatomic potentials. While on the mesoscale, the material entity is the bead, which consists of more than one atom, like a nanoparticle or grain \cite{nicolaides2001mesoscale}. Bead dynamics is governed by Newton’s equations of motion using molecular mechanical force fields to describe their interactions.  The materials entity at the microscale is the continuum volume, with the representative volume element (RVE) as the basic unit \cite{chen2024materials}\cite{blanco2016variational}. Several different governing physics equations are used (e.g. Newton’s equations, Navier-Stokes, Maxwell), depending on the physical properties that are studied.

Because of these very different ways of representing the materials and the different physics equations used at the three scales, independent scientific communities have developed, each with its own conceptualisations (ways to depict the world) and nomenclature tailored to describing the material and physical phenomena at the given scale. In order to bridge scales, the correct exchange of physical properties derived from the different representations of the material entities must be ensured \cite{raabe1998computational}\cite{choudhary2022recent}. Relying only on the name of the property will quickly lead to misunderstanding and errors.  For example, an interfacial energy calculated at the nanoscale using DFT may or may not include contributions from the strain field in the bulk structures, depending on how the model was set up and relaxed. To understand which of these choices best corresponds to what is called “interfacial energy” in a model at the microscale requires understanding that can be understood and correctly conceptualised  by the experts working at the nanoscale \cite{kuna2009effect}. With conceptualisation, we refer to the process of defining and correctly understanding the concepts (e.g. electron, atom, position, time) needed for expressing a scientific use case and how these concepts are related to each other. Conceptualisation is the preliminary step behind every scientific theory.

To share such conceptualisations and knowledge requires, first of all, a common language. It is important that this language is formal to avoid misunderstandings, but still understandable by humans and expressive enough to allow expressing the scientific knowledge and conceptualisations. Ontologies, which are languages formalised using logics, meet these requirements. The rise of the semantic web in the early 2000s has been adopted by information science and standardised by \href{https://www.w3.org/}{W3C} into the \href{https://www.w3.org/OWL/}{Web Ontology Language (OWL)}, which is a logic-based extension of the \href{https://www.w3.org/RDF/}{Resource Description Framework (RDF)} standard. Using OWL, one can express scientific knowledge in a concise way, verify its consistency and infer implicit knowledge.

\begin{figure}[H]
    \centering
    \includegraphics[width=0.8\linewidth]{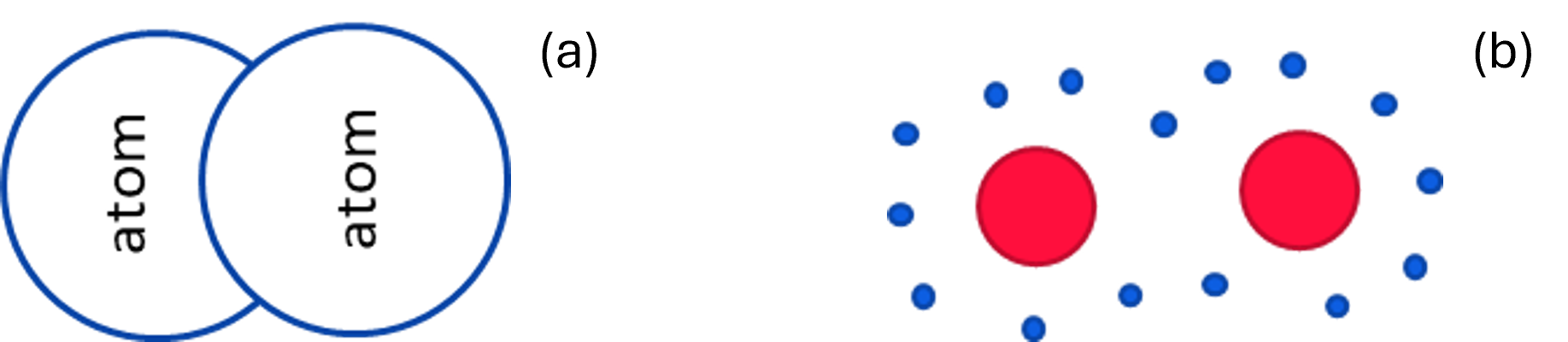}
    \caption{Incompatible definitions of a molecule in (a) chemistry and (b) physics.}
    \label{fig:molecule}
\end{figure}

Different scientific communities may have different conceptualisations of the world. An example is the definition of a molecule in the domains of chemistry and physics. According to \href{https://goldbook.iupac.org/}{IUPAC Goldbook} \cite{iupac2009goldbook}, which is an authoritative source within chemistry, a molecule is defined as an electrically neutral entity consisting of more than one atom, and an atom is a nucleus of Z positive charge and Z electrons. Physics lacks an authoritative source such as the IUPAC Goldbook, but a commonly agreed definition of a molecule is that it is an electrically neutral entity consisting of nuclei and electrons (a cloud of shared electrons). The incompatibility arises because the chemical definition presupposes atoms as well-defined entities with associated electrons, whereas the physical description treats electrons as fundamentally delocalised and does not require a unique atom–electron partition \cite{pauling1960nature}.  These seemingly incompatible definitions are illustrated in Figure~\ref{fig:molecule}. 

To be able to resolve issues like this, we must be able to share conceptualisations. For applied sciences, this requires an ontology build from the ground, based on scientific principles and knowledge. The \href{https://github.com/emmo-repo/EMMO}{Elementary Multidisciplinary Material Ontology (EMMO)} is an attempt to create such an ontology based on physics and applied sciences that can address challenges such as the abovementioned seemingly incompatible definitions. Its development has been led by the \href{https://emmc.eu/}{European Materials Modelling Council (EMMC)} since 2018. Its top and reference level has recently reached a stable version.  

\begin{figure}[H]
    \centering
    \includegraphics[width=0.5\linewidth]{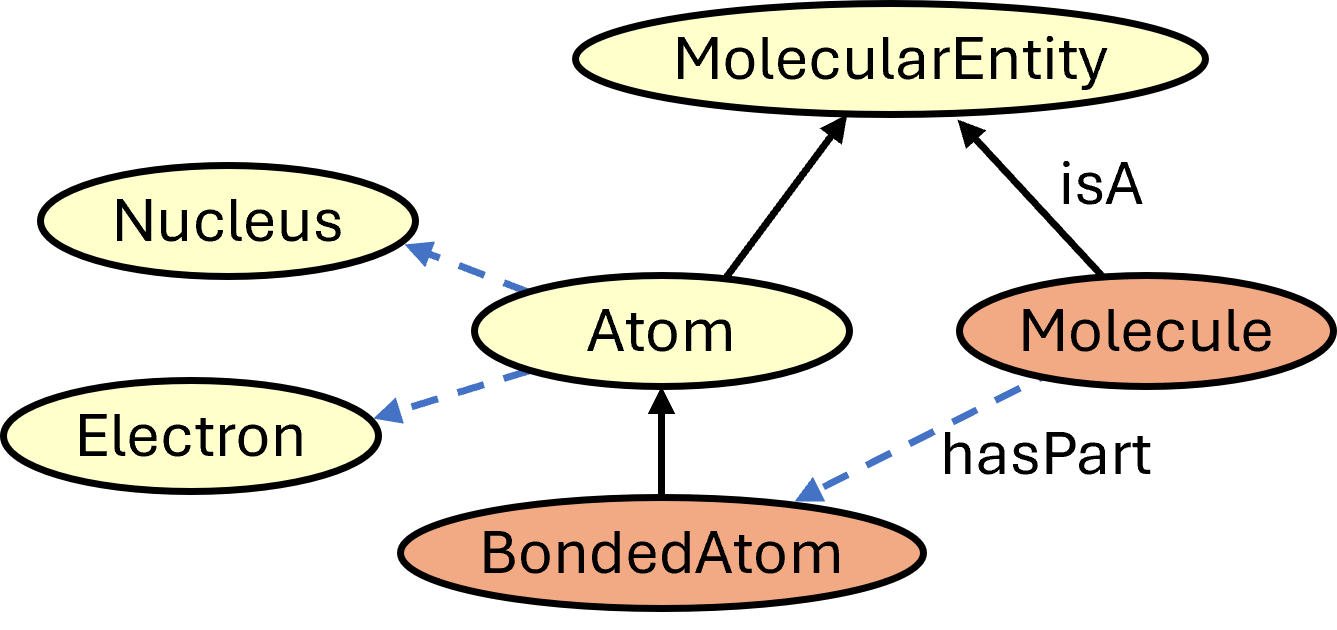}
    \caption{How EMMO aligns the different definitions of molecule and atom from the domains chemistry as defined in the IUPAC GoldBook (red) and physics (yellow). The concept of atom in chemistry has been labelled BondedAtom, which is a subclass of the atom as defined in physics. Likewise, the concept of molecule physics has been labelled MolecularEntity, which is a superclass of the molecule as defined in chemistry.}
    \label{fig:molecule-emmo}
\end{figure}

EMMO solves the seemingly incompatible definitions of a molecule by realising that the concepts referred to are not the same \cite{delnostro2024emmo}. EMMO defines the concepts of both domains and relate them to each other as illustrated in Figure~\ref{fig:molecule-emmo}.

\begin{figure}[H]
    \centering
    \includegraphics[width=0.8\linewidth]{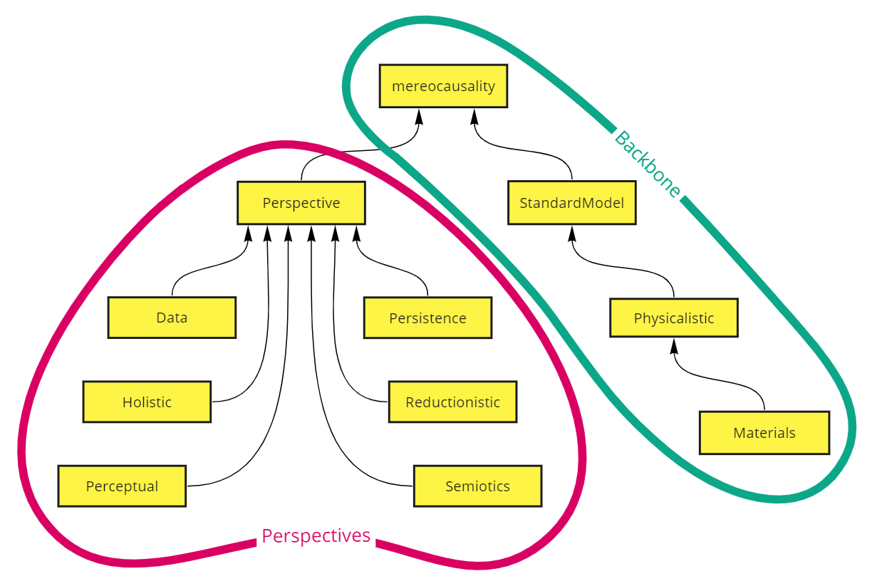}
    \caption{Illustration of the top-level modules of EMMO. At the very top, we find the mereocausality level, which introduces a fundamental theory combining mereology and causality. Based on this theory, EMMO provides two branches, perspective (red), which provides generic ways to categorise the world according to a perspective, and a backbone (green), which describes the world according to applied sciences, going from the standard model, via the fundamental entities of chemical-physics to the fundamental entities of materials science.}
    \label{fig:emmo-top}
\end{figure}

The structure of EMMO is shown in Figure~\ref{fig:emmo-top}. At the very top, we have the mereocausality module, which introduces a novel theory of mereocausality~\cite{zaccarini23} that combines the fundamental approaches of mereology~\cite{sep-mereology} and causality. Mereology is the theory about parthood and atomic mereology means that there exists a smallest part, the so-called \emph{Quantum}. Quanta may interact causally. Networks of such causally interacting quanta can be interpreted as macroscopic objects that themselves may interact either spatially or temporally, based on the topology of the underlying causal network. 

The \emph{standardmodel} module applies this theory to our knowledge of the real world, interpreting an elementary particle as a single chain of causally interacting quanta. This is extended by the \emph{physicalistic} module to more complex objects, such as atoms and molecules, that constitute the basic building blocks of physical chemistry. The \emph{materials} module further extends this and defines a taxonomy with the basic building blocks of materials science, like material, gas, alloy, crystal, etc. 

The \emph{perspectives} branch provides descriptions of the world seen from different perspectives. For example, entities are represented according to contrast, i.e. to variation in their properties, in the \emph{data} perspective. In the \emph{persistence} perspective, the world is categorised according to objects and processes and so on. A particularly important perspective is \emph{semiotics}, which is based on Pierce's semiotics~\cite{sep-peirce-semiotics}  and introduces the triadic semiotic process that relates an object to an interpreter and the produced sign. This theory provides the basis for defining properties as observables (sign) produced by the observer (interpreter) of the object, as is done in science, as opposed to intrinsic qualities of an object. The observer can be a measurement instrument or a computational procedure. 

This basis is extended by a \emph{multiperspective} level that combines the perspectives (like \emph{information}, which is data with meaning obtained by the combination of perspectives \emph{data} and \emph{semiosis}) and the \emph{discipline} level, with modules that provide common references within selected disciplines. Examples of disciplines include chemistry, metrology, models, and manufacturing. Based on this common core, the community has created a large set of domain ontologies, e.g. for characterisation methodologies, electrochemistry, chemical substances, microstructures, batteries, additive manufacturing, magnetic materials, concrete, etc. Together, these constitute a growing ontological ecosystem providing a common standardised reference for sharing knowledge between different scientific domains and across different scales. 

ML and AI tools benefit from being combined with ontological frameworks such as the one described above. The common reference enables interoperability between previously unconnected domains and the structure and context are essential for creating more intelligent and accurate ML and AI tools. Furthermore, the ontologies provide information on logically formulated relations between concepts, which gives AI and machines the possibility for reasoning and inferring new knowledge. ML and AI tools can analyse complicated relationships between data accurately and thus more effectively, thus reducing the chance of producing hallucinations.

\FloatBarrier

\section{The advent of large language models (LLMs)}

One of the biggest recent advances in ML has been the development of LLMs \cite{minaee2024large}. One of the most powerful aspects of LLMs is the simplicity of their functioning, interacting with a user by means of natural language prompts, rather than requiring complicated, specific computer programmes to obtain results. This simplicity of operation promises to lower the barrier to adaptation of LLMs and also provides a route to using LLMs to leverage other methods mentioned in this survey. Another striking feature of LLMs is how general their field of application appears to be. The majority of well-known LLMs are trained on a large corpous of text scraped from the internet. They are trained in an unsupervised or self-supervised manner, meaning that expensive annotation of the data (consistently identified as a barrier to other materials informatics approaches) is not an issue here \cite{devlin2019bert}. LLMs simply train by learning to predict missing tokens (words or segments of language). This deceptively simple training objective, when coupled with very large datasets, leads to models that perform surprisingly well in any number of disciplines.

\subsection{LLM Foundation models}

The generalisability of LLMs means that while they are able to perform surprisingly well on tasks that they have previously not seen, they often do not achieve state of the art performance on specific tasks. However, these "foundation" models can often be ‘fine-tuned’ to perform much more accurately on specific tasks \cite{jablonka2024leveraging}. Fine-tuning involves a partial re-training of the main model on a small dataset relevant to the domain of interest. In materials science, LLMs are adapted to comprehend and generate domain-specific language, facilitating tasks like information extraction, predictive modelling, and the generation of hypotheses, thereby bridging the gap between complex data and practical applications \cite{zhao2023beyond} \cite{ling2025domain}.

\begin{figure}[H]
\centering
\includegraphics[width=0.9\textwidth]{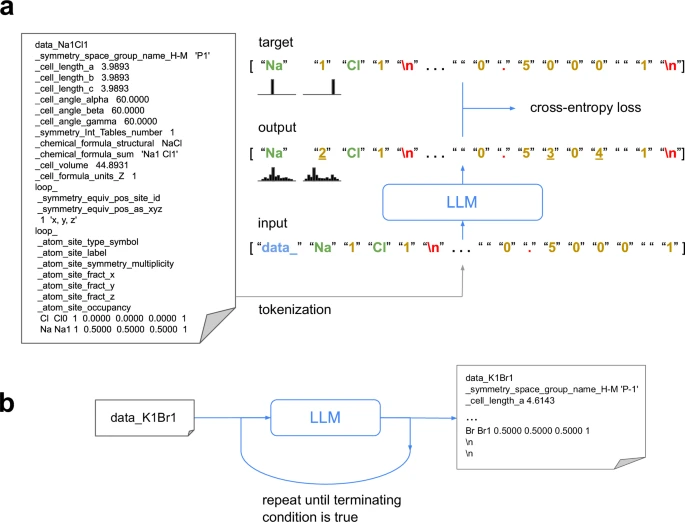}
\caption{The operating principle of an autoregressive LLM, in this case CrystaLLM. (a) Starting with a CIF file (left), the content is converted into a sequence of symbols through a process called tokenization. These tokens are then input into a machine learning model, which predicts the probability of each possible next token in the sequence. The model’s predictions are compared against the correct next tokens using a standard accuracy metric called cross-entropy loss. Here, the correct next token is defined as the one that actually follows in the original sequence. Tokens are color-coded based on their type: CIF tags (blue), atom symbols (green), numbers (gold), and punctuation (red). While training, the model doesn't generate actual text but learns to assign high probabilities to the correct next token. Tokens with underlines show examples where the model was less confident in its prediction.
(b) To generate a new CIF file, the process begins by feeding the model a starting prompt (e.g., "data" followed by the desired chemical composition). The model then predicts the next token, which is sampled and added to the output. This prediction–sampling cycle repeats, building the CIF file token by token, until a stopping condition is reached such as generating two newline characters in a row. Reproduced with
permission from Ref. \cite{antunes2024crystal}.}
 \label{fig:crystallm.png}
\end{figure}

Foundation models have been applied in materials science for tasks such as property prediction, synthesis planning and molecular generation. For example, the Molecular Transformer model \cite{schwaller2019molecular} treated reaction pathways as a language translation problem and demonstrated one of the first applications of LLMs to reaction design.
Furthermore, integration of LLMs with natural language processing techniques is enhancing how we interact with the scientific literature. Scientifically aware LLMs can be sued to extract important information from large corpora of text, assisting in the building of datasets to be used in other materials informatics settings. LLMs can also be used to summarise large numbers of papers and provide researchers with an overview of a collection of sources. For istance, CrystaLLM is a decoder-only, Transformer-based language model designed to model the Crystallographic Information File (CIF) format. We have recently trained CrystaLLM autoregressively\cite{antunes2024crystal} on a large-scale corpus comprising millions of CIF files as shown in Figure ~\ref{fig:crystallm.png}a. Unlike approaches that rely on structured representations extracted from CIFs, CrystaLLM operates directly on the standardized, tokenized textual content of the CIF files. During training, the model is presented with sequences of CIF tokens and learns to predict subsequent tokens in an autoregressive manner. After training, the model can generate novel CIF files by conditioning on an initial sequence of tokens and iteratively sampling subsequent tokens based on the previously generated content. This process continues until a predefined termination criterion is met, resulting in a complete CIF file (Figure ~\ref{fig:crystallm.png}b). 

\subsection{Generative modelling with LLMs}

Generative modelling is a class of ML that is designed to generate new data that resembles the data from a distribution upon which the model has been trained. Unlike methods that predict properties or classify systems, generative models aim to learn underlying patterns in complex datasets and use these patterns to hypothesise completely new examples – these methods are used in many fields from images to text to molecules and crystals. In the context of materials science, this means generating new crystal structures or molecules that are physically plausible. This ability to generate new materials is promising for the acceleration of discoveries. Traditional approaches based on human intuition coupled with high-throughput experimentation and/or computation are very expensive and time-consuming. Generative models have the potential to shift this paradigm by proposing new systems that satisfy certain design constraints, thereby focusing the experimental search space.

\subsection{LLM Agents}

AI agents are systems that combine LLMs with external tools, structured reasoning capabilities, and memory to autonomously perform complex tasks. Unlike static LLMs that generate one-off responses, agents operate iteratively, planning steps, invoking tools (e.g., databases, simulation software), and learning from feedback. In chemistry and materials science, this means going beyond simple prediction or text generation to carry out research-like workflows: literature searches, hypothesis generation, experimental design, and even interfacing with laboratory robots. Materials discovery is an iterative, multi-step process involving knowledge synthesis, candidate generation, simulation or experiment, and interpretation. AI agents promise to automate parts – or even all – of this pipeline. Their ability to integrate domain-specific tools and contextual memory means they could act as ``autonomous research assistants, speeding up the pace of discovery and enabling more efficient exploration of vast chemical design spaces.

\begin{figure}[H]
\centering
\includegraphics[width=0.9\textwidth]{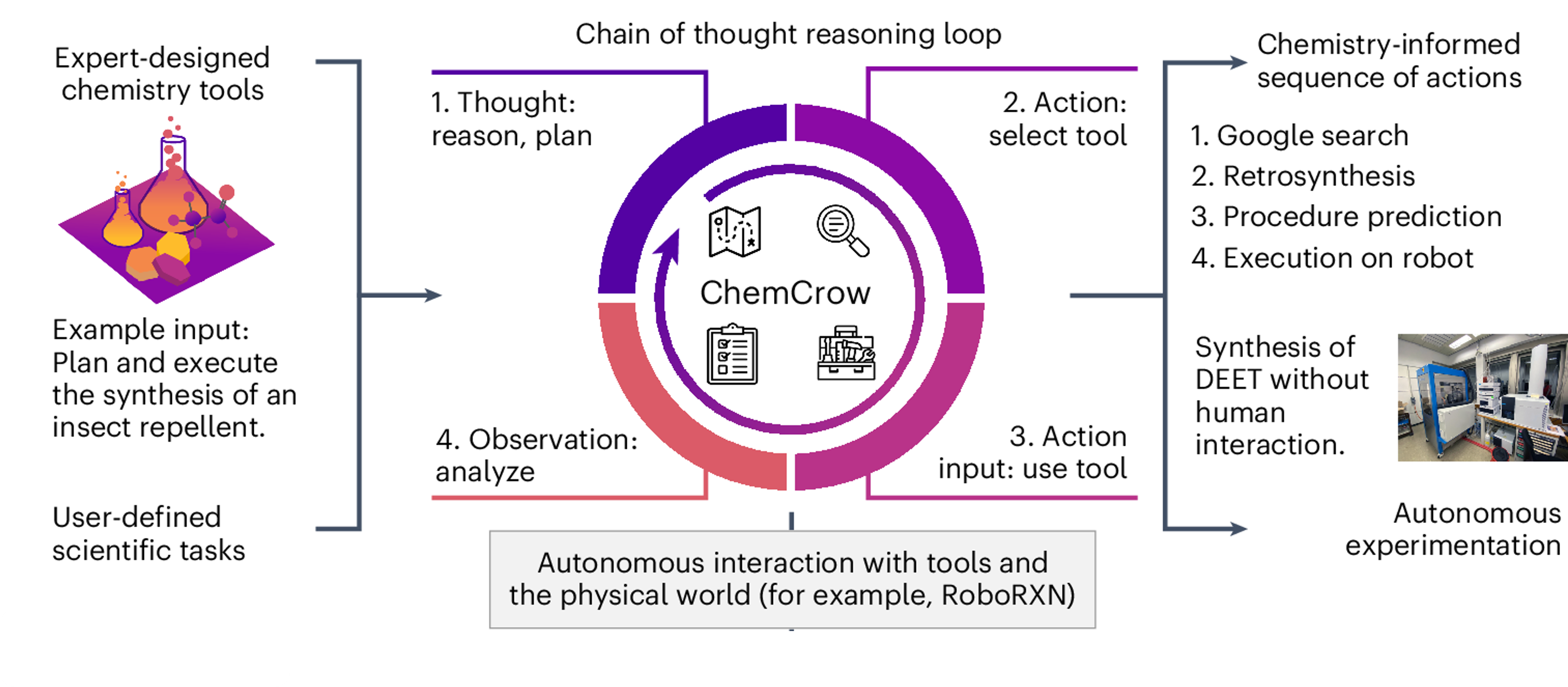}
\caption[LLM agent system solving a task]{An illustration of how an LLM agent system solves a task. A collection of chemistry-related software tools is assembled and made available to a large language model (LLM). When given a user query, the LLM uses these tools in an automatic, step-by-step reasoning process, choosing which tools to use, what inputs to provide, and how to proceed until it reaches a final answer. In this example, the system is tasked with planning the synthesis of DEET, a widely used insect repellent. Reproduced with permission from Ref. \cite{m2024augmenting}.}
\label{fig:Figure_9.png}
\end{figure}

For example, Inspired by successful applications of large language models in other scientific domains \cite{schick2023toolformer} \cite{yang2023mm}, we have introduced ChemCrow, an LLM-powered chemistry engine designed to streamline reasoning and decision-making across a wide range of chemical tasks, including drug discovery, materials design, and chemical synthesis \cite{m2024augmenting}. ChemCrow integrates a suite of expert-designed chemistry tools and leverages a large language model (GPT-4 in our experiments) to coordinate their use, as illustrated in Figure ~\ref{fig:Figure_9.png}. The LLM is prompted with explicit instructions describing the task, the desired output format, and a catalog of available tools, including their functionality and expected input–output specifications. Given a user-defined query, the model determines when and how to invoke these tools to effectively address the task. To structure its reasoning and actions, ChemCrow follows the Thought–Action–Action Input–Observation framework \cite{yao2022react}. In the \textit{Thought} phase, the model analyzes the current state of thanalysesm, assesses its relevance to the overall objective, and plans the next steps. It then selects an appropriate tool (\textit{Action}) and provides the required inputs (\textit{Action Input}).

At this point, text generation is paused while the requested tool is executed programmatically. The resulting output is returned to the model as an \textit{Observation}, after which the LLM re-enters the \textit{Thought} phase. This closed-loop process repeats iteratively, enabling adaptive reasoning and tool usage, until the final solution is produced.

LLMs were augmented with chemistry toolkits such as RDKit, PubChem queries, and structure editors through a technique known as “tool augmentation". Here, LLMs act as the reasoning engine while delegating specialised tasks like drawing a molecule or checking for stability to external tools. This dramatically improves the model’s accuracy and reliability in real-world chemical questions, compared to using a standalone LLM \cite{bran2023chemcrow}. Building on this, researchers have constructed fully-fledged autonomous agents by wrapping LLMs in structured environments that enable tool use, planning, memory, and feedback loops. These agents can autonomously search literature, extract chemical knowledge, run simulations, and even plan retrosynthetic routes. For example, some systems have been benchmarked on tasks like reagent lookup, functional group identification, or property estimation, achieving success rates far beyond static prompting methods \cite{ramos2025review}.

As large language models such as LLMs and other machine learning systems become increasingly central to materials science, robust and meaningful evaluation frameworks are essential to ensure genuine progress. As recently highlighted \cite{alampara2025lessons}, current evaluation practices, and poorly designed benchmarks can lead to "phantom progress" - where models appear to improve on metrics but fail in real-world applications. Drawing from statistical measurement theory, the authors argue that evaluations must go beyond ease of implementation and focus on construct validity: are we truly measuring the capabilities that matter? Key challenges include dataset biases, oversimplified metrics, poor reproducibility, and a lack of transparency. There is an overarching need for diverse, representative test sets; we would caution against the overuse of “solved” benchmarks like QM9.

\section{Conclusions}
In this review, we provided a comprehensive survey of materials informatics methods across length scales, from the atomic and electronic scale to the continuum level. Our survey highlights how multi-scale modelling and data-driven techniques are transforming the way materials are discovered, characterised, and optimised. At the electronic, atomistic level and at the nanoscale, simulation techniques range from ab initio methods, such as DFT, to large-scale particle-based molecular dynamics based on empirical force fields. These approaches enable accurate predictions of fundamental electronic, dynamical and structural properties of materials, forming the foundational layer of multiscale models. ML techniques at this scale are being used not only to accelerate quantum simulations but also to generalise them, predicting properties for vast chemical and structural spaces that would otherwise be computationally intractable. Also, these models provide inputs for mesoscale and macroscale simulations, highlighting the need for vertical integration across scales. At the mesoscale and microscale, materials informatics tools can describe the properties of materials related, for example, to grain structures, phase boundaries, and mechanical behaviour. In this context, experimental knowledge can successfully be integrated into the simulation workflow to enable a comprehensive data-driven approach to materials investigation. The rise of deep learning in image analysis, the use of graph neural networks for microstructure-property relationships, and reinforcement learning in process optimisation all exemplify the growing sophistication of these tools.

However, despite progress at individual scales, one of the most pressing open challenges is the integration of these approaches into coherent multiscale frameworks. Most existing tools and models remain tailored to a specific length or time scale, making interoperability difficult. To bridge this gap, it is essential to move toward data-driven multiscale integration, developing standardised data formats, shared feature representations, and physics-informed ML models capable of transferring knowledge across scales. Such integration would allow, for example, electronic-scale predictions to inform mesoscale simulations or continuum-level models. The advent of LLMs further supports this vision, offering tools to navigate and connect knowledge across disciplines, extract structured data from the literature, and even generate interoperable workflows that bridge atomistic, mesoscopic, and macroscopic perspectives. These capabilities can help address the current fragmentation of efforts across scales, fostering truly integrated modelling pipelines.
Moreover, none of these advances can be fully realised in isolation. The path forward must be paved with collaborative, community-driven initiatives that enable resource sharing, joint development of interoperable platforms, and the adoption of FAIR data principles. Networking activities, such as consortia, open working groups, benchmarking challenges, and open workshops, are essential to build a common language across communities working at different scales.

Therefore, the future of materials informatics lies not just in better models or larger datasets, but in a unified, scale-aware, data-centric ecosystem. Such an ecosystem must integrate structure and property descriptors across scales through shared data infrastructures and interoperable AI tools. By supporting cross-scale learning and interdisciplinary collaboration, the materials informatics community can accelerate the transition from fragmented workflows to holistic materials design strategies, unlocking new classes of high-performance, sustainable materials for innovation in science and technology.

\section{Acknowledgements}
This article is based upon work from COST Action EuMINe - European Materials Informatics Network (CA22143), supported by COST  (European Cooperation in Science and Technology).
Partial funding is also acknowledged from the Horizon Europe PINK (GA no. 101137809) and MatCHMaker (GA no. 101091687) projects as well as the Norwegian SFI PhysMet (RCN 309584) project.
KTB acknowledges support from UKRI (EP/Y000552/1 and EP/Y014405/1).

\bibliographystyle{unsrt}
\bibliography{references}

\end{document}